\documentstyle[fleqn]{article}
\begin{document}
\LARGE
\begin{center}
On an Intriguing Signature-Reversal Exhibited by\\ Cremonian
Spacetimes
\end{center}
\vspace*{.2cm}
\Large
\begin{center}
Metod Saniga

\vspace*{.3cm} \small {\it Astronomical Institute, Slovak Academy
of Sciences, 05960 Tatransk\' a Lomnica, Slovak Republic}
\end{center}

\vspace*{-.4cm}
\noindent
\hrulefill

\vspace*{.2cm}
\small
\noindent
{\bf Abstract}\\

\noindent It is shown that a generic quadro-quartic Cremonian
spacetime, which is endowed with one spatial and three time
dimensions, can {\it continuously} evolve into a signature-{\it
reversed} configuration, i.e. into the classical spacetime
featuring one temporal and three space dimensions. An interesting
cosmological implication of this finding is mentioned.

\noindent \hrulefill

\vspace*{.5cm} \normalsize \noindent In a recent series of papers
[1--4], we have introduced and examined in detail the concept of
so-called Cremonian spacetimes. A Cremonian spacetime is defined
as an algebraic geometrical configuration that lives in a
three-dimensional projective space and consists of so-called
fundamental elements of a Cremona transformation generated by a
homaloidal web of ruled quadrics. The fundamental elements are
found to be of {\it two} different kinds, viz. lines and conics,
and form distinct {\it pencils}, i.e. linear, single parametrical
aggregates. These pencils are taken to represent macroscopic
dimensions of the real physical world: those comprising lines --
space, those consisting of conics -- time. We have further
demonstrated that all generic Cremonian spacetimes which are
associated with quadric-generated Cremona transformations whose
inverses are generated by cubic or quartic surfaces are endowed
with four dimensions. Yet, their signatures are not all the same.
While the quadro-{\it cubic} spacetimes [1,2] are endowed with
three spatial and one time dimensions (and being thus compatible
with what Nature offers to our senses), the quadro-{\it quartic}
ones [3] feature just the opposite, i.e. three temporal dimensions
and a single spatial one. The aim of this note is to show that
there exists a continuous deformation of a generic quadro-quartic
spacetime resulting in a signature-{\it reversed} manifold; the
latter being a specific, `spatially anisotropic' quadro-cubic
Cremonian spacetime [2].

With this end in view, we shall consider, following the symbols
and notation of [1--3], a homaloidal web of quadrics
\begin{equation}
{\cal W}_{\vartheta}^{\kappa}:~~
\vartheta_{1}\breve{z}_{3}(\breve{z}_{2} - \breve{z}_{3}) +
\vartheta_{2} \breve{z}_{3}(\breve{z}_{1} - \breve{z}_{3}) +
\vartheta_{3} (\breve{z}_{1}\breve{z}_{2} - \breve{z}_{3}^{2})+
\vartheta_{4} \breve{z}_{4}(\kappa_{1} \breve{z}_{1} -
\kappa_{2}\breve{z}_{2}) = 0,
\end{equation}
with $\kappa_{1}, \kappa_{2}$ being real-valued parameters. The
web generates the following Cremona transformation from one
(`unprimed') projective space, $P_{3}$, into other (`primed')
projective space, $P'_{3}$,
\begin{equation}
\varrho \breve{z}'_{1} = \breve{z}_{3}(\breve{z}_{2} -
\breve{z}_{3}),~~ \varrho \breve{z}'_{2} =
\breve{z}_{3}(\breve{z}_{1} - \breve{z}_{3}),~~ \varrho
\breve{z}'_{3} = (\breve{z}_{1}\breve{z}_{2} -
\breve{z}_{3}^{2}),~~ \varrho \breve{z}'_{4} =
\breve{z}_{4}(\kappa_{1} \breve{z}_{1} - \kappa_{2}\breve{z}_{2}),
\end{equation}
where $\varrho$ is a non-zero proportionality factor. Our task is
to find the structure of the configuration of fundamental elements
associated with the above transformation. To furnish this task, we
recall [1] that a fundamental (also known, especially in older
literature [5], as principal) element associated with a given
Cremona transformation is the totality of points, either a curve
or a surface, in the first space whose counterpart (homologue) in
the other space is just a {\it single} point. Upon combining Eqs.
(1) and (2) we find that in the case under consideration the
fundamental elements are constituents of the following {\it four}
pencils:
\begin{equation}
{\cal A}^{\kappa}_{\vartheta}:~~  \breve{z}_{2} -\breve{z}_{3} = 0
= \breve{z}_{3}(\breve{z}_{1} -\breve{z}_{3}) + \vartheta
\breve{z}_{4} (\kappa_{1} \breve{z}_{1} - \kappa_{2}
\breve{z}_{3}), ~~ \vartheta \equiv \vartheta_{4}/(\vartheta_{2} +
\vartheta_{3}),
\end{equation}
\begin{equation}
{\cal B}^{\kappa}_{\vartheta}:~~  \breve{z}_{1} -\breve{z}_{3} = 0
= \breve{z}_{3}(\breve{z}_{2} -\breve{z}_{3}) + \vartheta
\breve{z}_{4} (\kappa_{1} \breve{z}_{3} - \kappa_{2}
\breve{z}_{2}), ~~ \vartheta \equiv \vartheta_{4}/(\vartheta_{1} +
\vartheta_{3}),
\end{equation}
\begin{equation}
{\cal C}^{\kappa}_{\vartheta}:~~  \breve{z}_{3} = 0 =
\breve{z}_{1} \breve{z}_{2} + \vartheta \breve{z}_{4} (\kappa_{1}
\breve{z}_{1} - \kappa_{2} \breve{z}_{2}), ~~ \vartheta \equiv
\vartheta_{4}/\vartheta_{3},
\end{equation}
and
\begin{equation}
{\cal D}^{\kappa}_{\vartheta}:~~ \kappa_{1} \breve{z}_{1} -
\kappa_{2} \breve{z}_{2} = 0 = \vartheta_{3} \kappa_{1}^{2}
\breve{z}_{1}^{2} + \left( \vartheta_{1} \kappa_{1}^{2} +
\vartheta_{2} \kappa_{1} \kappa_{2} \right)
\breve{z}_{1}\breve{z}_{3} - \kappa_{1}\kappa_{2}
(\vartheta_{1}+\vartheta_{2}+\vartheta_{3}) \breve{z}_{3}^{2},
\end{equation}
which can equivalently be written as
\begin{equation}
{\cal D}^{\kappa}_{\vartheta}:~~ \kappa_{1} \breve{z}_{1} -
\kappa_{2} \breve{z}_{2} = 0 = \vartheta_{3} \kappa_{2}^{2}
\breve{z}_{2}^{2} + \left( \vartheta_{1} \kappa_{1}\kappa_{2}  +
\vartheta_{2} \kappa_{2}^{2} \right) \breve{z}_{2}\breve{z}_{3} -
\kappa_{1}\kappa_{2} (\vartheta_{1}+\vartheta_{2}+\vartheta_{3})
\breve{z}_{3}^{2}.
\end{equation}
And now we arrive at a crucial observation: whereas the last
pencil comprises one and the same kind of geometrical objects,
namely {\it lines}, irrespective of the value of the parameters
$\kappa_{1}, \kappa_{2}$, this is not the case with the other
three pencils! Although, in general, each of these consists of
proper {\it conics}, there exist particular values of $\kappa_{1}$
and $\kappa_{2}$ when the conics of a given pencil become {\it
all} composite, featuring pairs of distinct real, coincident real,
or conjugate complex {\it lines}. It represents no difficulty to
find out when this happens. We shall first consider the ${\cal
A}$-pencil. After introducing a more convenient, `affine'
parameter $\kappa \equiv \kappa_{2}/\kappa_{1}$, we see that this
pencil consists of lines if and only if $\kappa = 1$, viz.
\begin{equation}
{\cal A}^{\kappa = 1}_{\vartheta}:~~  \breve{z}_{2} -\breve{z}_{3}
= 0 = \breve{z}_{3} + \vartheta g \breve{z}_{4},~~(g \equiv
\kappa_{1}= \kappa_{2})
\end{equation}
or $\kappa = \infty$, viz.
\begin{equation}
{\cal A}^{\kappa = \infty}_{\vartheta}:~~  \breve{z}_{2}
-\breve{z}_{3} = 0 = \breve{z}_{1} - \breve{z}_{3} - \vartheta
\kappa_{2} \breve{z}_{4},
\end{equation}
where $\vartheta \equiv \vartheta_{4}/(\vartheta_{2} +
\vartheta_{3})$. Similarly, the ${\cal B}$-pencil features lines
for $\kappa = 0$, viz.
\begin{equation}
{\cal B}^{\kappa = 0}_{\vartheta}:~~  \breve{z}_{1} -\breve{z}_{3}
= 0 = \breve{z}_{2} - \breve{z}_{3} + \vartheta \kappa_{1}
\breve{z}_{4},
\end{equation}
and $\kappa = 1$, viz.
\begin{equation}
{\cal B}^{\kappa = 1}_{\vartheta}:~~  \breve{z}_{1} -
\breve{z}_{3} = 0 = \breve{z}_{3} - \vartheta g \breve{z}_{4},
\end{equation}
where $\vartheta \equiv \vartheta_{4}/(\vartheta_{1} +
\vartheta_{3})$. Finally, the ${\cal C}$-pencil is found to
comprise lines for $\kappa = 0$, viz.
\begin{equation}
{\cal C}^{\kappa = 0}_{\vartheta}:~~  \breve{z}_{3} = 0 =
\breve{z}_{2} + \vartheta \kappa_{1} \breve{z}_{4},
\end{equation}
and $\kappa = \infty$, viz.
\begin{equation}
{\cal C}^{\kappa = \infty}_{\vartheta}:~~  \breve{z}_{3} = 0 =
\breve{z}_{1} - \vartheta \kappa_{2} \breve{z}_{4},
\end{equation}
where $\vartheta \equiv \vartheta_{4}/\vartheta_{3}$. Our findings
can succinctly be summarized as follows:

\vspace*{.1cm}
\begin{center}
\begin{tabular}{||c||c|c|c|c||}
\hline \hline $\kappa$ & 0 & 1 & $\infty$ & other \\
\hline \hline ${\cal A}^{\kappa}$ & conics & lines & lines & conics \\
\hline ~${\cal B}^{\kappa}$~ & lines & lines & conics & conics \\
\hline ${\cal C}^{\kappa}$ & lines & conics & lines & conics \\
\hline ${\cal D}^{\kappa}$ & lines & lines & lines & lines \\
\hline \hline
\end{tabular}
\end{center}

\vspace*{.1cm}

There are several important features readily discernible from this
table. First, there exists a (just recently discovered [1--3],)
totally amazing {\it three-to-one} splitting in the character of
the pencils of fundamental elements {\it regardless} of the value
of $\kappa$; that is, {\it one} of the pencils is {\it always} of
a qualitatively different nature than the remaining {\it three}.
Second, the far prevailing mode is the 1+3 one, i.e. the
configuration (Cremonian spacetime) with {\it one} pencil of {\it
lines} (one spatial dimension) and {\it three} pencils of {\it
conics} (three time dimensions); the three 3+1 configurations can
be seen as mere `islands' in the `sea' of 1+3's. The third, and
perhaps most intriguing, fact is that we can freely move on the
$\kappa$-axis in such a way that wherever we start we can always
reach one of the islands; in other words, a continuous variation
of the parameter $\kappa$ must always be accompanied by gradual
qualitative changes in the structure of the initial 1+3
configuration so that this configuration will eventually be
transformed into a 3+1 manifold.

In order to get a deeper insight into the nature of this
`signature-reversal' phenomenon, we shall have a closer look at
the {\it base} (i.e. shared by {\it all} the quadrics) elements of
our homaloidal web, Eq. (1). It is easy to verify that if $\kappa$
differs from 0, 1 and $\infty$ the only base elements are four
distinct, non-coplanar points, namely ($\varrho \neq 0$)
\begin{equation}
{\rm B}_{1}:~~ \varrho \breve{z}_{\alpha} = (0, 1, 0, 0),
\end{equation}
\begin{equation}
{\rm B}_{2}:~~ \varrho \breve{z}_{\alpha} = (1, 1, 1, 0),
\end{equation}
\begin{equation}
{\rm B}_{3}:~~ \varrho \breve{z}_{\alpha} = (1, 0, 0, 0),
\end{equation}
and
\begin{equation}
{\rm B}:~~ \varrho \breve{z}_{\alpha} = (0, 0, 0, 1);
\end{equation}
this means that the corresponding Cremona transformation, Eq. (2),
is of a so-called quadro-{\it quartic} type [3,5]. If, on the
other hand, $\kappa$ = 0, 1, and/or $\infty$ the web is endowed,
in addition to the four points, with a whole {\it line} of base
points, namely the $\breve{z}_{1} = 0 = \breve{z}_{3}$,
$\breve{z}_{1} - \breve{z}_{2} = 0 = \breve{z}_{2} -
\breve{z}_{3}$, and/or $\breve{z}_{2} = 0 = \breve{z}_{3}$ one,
respectively; in this case, the corresponding Cremona
transformation is of a qualitatively different, so-called
quadro-{\it cubic} type [1,2,5]. Moreover, this particular
transformation is not most general because the base line, which
necessarily passes through the point B, contains in each case also
one of the remaining three base points, namely the point B$_{1}$,
B$_{2}$ and/or B$_{3}$, respectively. Hence, the corresponding
Cremonian spacetime, first studied and described in detail in [2],
is found to be a little peculiar: it displays a sort of `spatial
anisotropy' in the sense that one of its space dimensions has a
slightly different footing than the other two.

The findings of this paper may obviously turn out to be of some
relevance to cosmology. There is a growing suspicion among
physicists [e.g. 6--10] that the Universe might have been born
with a different (macro-)signature, and even a different
(macro-)dimensionality, than we currently observe. It may well be
that the `original' signature was just the opposite, i.e. that the
Universe was created with a single space and three time
dimensions, and the current signature could simply be a result of
the above-outlined Cremonian evolutionary `jump.' If this scenario
is correct then future more sophisticated astrophysical
observations are bound to reveal, as already stipulated in [2],
that one of the three macroscopic space dimensions is slightly at
odds with the other two. This feature would not only pose a
serious challenge to some of the currently favoured physical
paradigms (like, e.g., CPT invariance), but would also raise a
host of profound epistemological and ontological questions.

\vspace*{.4cm} \noindent \small
{\bf Acknowledgements}\\
This work was supported in part by a 2001/2002 NATO/FNRS Advanced
Research Fellowship and the 2000--2002 NATO Collaborative Linkage
Grant PST.CLG.976850. I am very grateful to Avshalom Elitzur and
Rosolino Buccheri for their constructive comments/remarks on the
first draft of the paper.

\end{document}